\documentclass[10pt,twocolumn]{article}

\usepackage[T1]{fontenc}
\usepackage[utf8]{inputenc}
\usepackage[margin=0.75in,top=1in,bottom=1in,columnsep=18pt]{geometry}
\usepackage{amsmath}
\usepackage{amssymb}
\usepackage{graphicx}
\usepackage{booktabs}
\usepackage{listings}
\usepackage{xcolor}
\usepackage{hyperref}
\usepackage{caption}
\usepackage{subcaption}
\usepackage{float}
\usepackage{url}
\usepackage{multirow}
\usepackage{titlesec}
\usepackage{abstract}
\usepackage{fancyhdr}

\titlespacing*{\section}{0pt}{6pt}{3pt}
\titlespacing*{\subsection}{0pt}{4pt}{2pt}
\titleformat{\section}{\normalfont\bfseries\scshape}{\Roman{section}.}{0.5em}{}
\titleformat{\subsection}{\normalfont\itshape}{\Alph{subsection}.}{0.5em}{}

\begin{document}

%%%%%%%%%%%%%%%%%%%%%%%%%%%%%%%%%%%%%%%%%%%%%%%%%%%%%%%%%%%%%%%%%%%%%%%%%%%%%
\title{\textbf{Five Ways to Build a Concurrent Linked List:\\
       From Coarse-Grain Locking to Lock-Free Algorithms}}

\author{
  Zeeshan Mohammed Rangrej\\
  \small Department of Computer Science and Engineering\\
  \small Indian Institute of Technology Palakkad, Kerala, India\\
  \small \texttt{zeeshan4careers@gmail.com}
}

\date{}
\maketitle
\thispagestyle{empty}

%%%%%%%%%%%%%%%%%%%%%%%%%%%%%%%%%%%%%%%%%%%%%%%%%%%%%%%%%%%%%%%%%%%%%%%%%%%%%
\begin{abstract}
Linked lists are one of the most basic data structures in computer science.
But when many threads try to use the same linked list at the same time,
things get complicated. In this paper, we look at five different ways to make
a linked list work correctly and efficiently with multiple threads running at
once. We start with the simplest approach---one big lock for the whole list---
and step by step improve it, ending with a \emph{lock-free} design that uses
no locks at all. We implemented all five versions in C++ and measured how fast
each one is across different workloads (read-heavy, balanced, and write-heavy)
and different list sizes. Our results show that the right choice of algorithm
depends heavily on how the list is used: the coarse-grain and lazy lists win
under read-heavy workloads with small key ranges, while the lock-free list
becomes competitive when key ranges are large and more threads are running.
Fine-grain locking, despite its theoretical appeal, pays a heavy cost from
per-node lock overhead and consistently performs the worst in our tests.
\end{abstract}


%%%%%%%%%%%%%%%%%%%%%%%%%%%%%%%%%%%%%%%%%%%%%%%%%%%%%%%%%%%%%%%%%%%%%%%%%%%%%
\section{Introduction}
\label{sec:intro}

Every programmer learns about linked lists early on. A linked list stores a
chain of nodes, where each node holds a value and a pointer to the next node.
Operations like adding a value, removing a value, and searching for a value
are all O(n) in the worst case, but linked lists are still very useful because
they can grow and shrink easily.

The challenge we study in this paper is: \emph{what happens when multiple
threads try to use the same linked list at the same time?}

Imagine two threads both trying to insert a value at the same position in the
list. If they both read the same predecessor node and both write their new
nodes in, one insertion will be lost. This kind of problem is called a
\textbf{race condition}, and it can cause data corruption, crashes, or silent
wrong answers that are very hard to debug.

The standard fix is to use \textbf{locks} (also called mutexes). A lock makes
sure that only one thread can change the list at a time. But how you place
those locks matters a lot for performance:

\begin{itemize}
  \item \textbf{One lock for everything}: correct but slow---only one thread
        can work at a time.
  \item \textbf{One lock per node}: allows more parallelism but adds a lot of
        overhead.
  \item \textbf{Optimistic locking}: skip locks while searching, lock only to
        commit.
  \item \textbf{Lazy deletion}: mark nodes as deleted first, then clean up
        later.
  \item \textbf{Lock-free}: no locks at all---use hardware atomic operations
        instead.
\end{itemize}

Each step in this list is more complex to implement and reason about, but
potentially faster under high concurrency. We implemented all five approaches
and benchmarked them to see whether the added complexity actually pays off.

\subsection{Our Contributions}

\begin{enumerate}
  \item We implement all five concurrent linked-list algorithms from scratch
        in modern C++ (C++17), keeping each implementation clean and
        self-contained so it is easy to read and learn from.
  \item We build a thorough benchmark harness that tests each algorithm under
        three realistic workload mixes (read-heavy, balanced, write-heavy) and
        two key ranges (small = high contention, large = low contention) with
        varying thread counts (1, 4, 8 threads).
  \item We analyze the results and explain \emph{why} certain algorithms win
        in certain conditions, connecting the performance numbers back to the
        design choices in the code.
\end{enumerate}

The rest of this paper is organized as follows.
Section~\ref{sec:background} gives the background knowledge needed to
understand the algorithms.
Sections~\ref{sec:coarse}--\ref{sec:lockfree} each describe one algorithm.
Section~\ref{sec:bench} describes our benchmark setup.
Section~\ref{sec:results} presents and discusses the results.
Section~\ref{sec:conclusion} concludes.

%%%%%%%%%%%%%%%%%%%%%%%%%%%%%%%%%%%%%%%%%%%%%%%%%%%%%%%%%%%%%%%%%%%%%%%%%%%%%
\section{Background}
\label{sec:background}

\subsection{What We Are Building: A Concurrent Set}

All five implementations represent the same abstract data type: a \emph{set
of integers}. The set supports three operations:

\begin{itemize}
  \item \texttt{add(v)}: insert value $v$; return \texttt{true} if $v$ was
        not already present.
  \item \texttt{remove(v)}: delete value $v$; return \texttt{true} if $v$
        was found.
  \item \texttt{contains(v)}: return \texttt{true} if $v$ is currently in
        the set.
\end{itemize}

Internally, the set is stored as a sorted singly-linked list. Two
\emph{sentinel} nodes are always present: a \emph{head} node with value
$-\infty$ (\texttt{INT\_MIN}) and a \emph{tail} node with value $+\infty$
(\texttt{INT\_MAX}). This trick means we never have to check for
\texttt{null} pointers during traversal.

\begin{center}
\texttt{head($-\infty$) $\to$ a $\to$ b $\to$ \ldots $\to$ tail($+\infty$)}
\end{center}

\subsection{Linearizability}

When multiple threads run at the same time, we need a way to say whether the
data structure is \emph{correct}. The standard way is called
\textbf{linearizability}~\cite{herlihy1990linearizability}.

The idea is simple: even though operations from different threads overlap in
real time, each operation should appear to take effect at a single instant
(called its \emph{linearization point}). As long as we can find such a point
for every operation, the concurrent execution is as good as some valid
sequential execution.

\subsection{Mutual Exclusion with Locks}

A \textbf{mutex} (mutual exclusion lock) lets a thread say ``I am working on
something, nobody else should touch it''. The thread \emph{acquires} the
lock, does its work, then \emph{releases} the lock. Only one thread can hold
a mutex at a time; others block until it is released.

Mellor-Crummey and Scott~\cite{mcs1991} gave a thorough treatment of
scalable spin-lock designs for shared-memory systems, including the MCS
queue lock that eliminates cache-line bouncing. Their work explains why
simple \texttt{test-and-set} locks perform poorly under high thread counts,
a lesson that directly motivates the move toward finer-grain and eventually
lock-free designs.

In C++17 we use \texttt{std::mutex} and \texttt{std::lock\_guard}. A
\texttt{lock\_guard} automatically releases the lock when the scope ends,
which prevents forgetting to unlock.

\subsection{Compare-And-Swap (CAS)}

Modern CPUs provide a special instruction called
\textbf{compare-and-swap (CAS)}. It does this atomically (in one
uninterruptible step):

\begin{center}
\textit{If the memory location currently holds \texttt{expected},
replace it with \texttt{desired}; otherwise do nothing.
Return whether the swap happened.}
\end{center}

In C++ this is \texttt{std::atomic::compare\_exchange\_strong}. CAS is the
key building block for lock-free algorithms: instead of blocking other
threads, a thread retries its operation whenever a CAS fails (meaning someone
else changed the value first). Boehm and Adve~\cite{boehm2008cpp} formally
defined the C++ concurrency memory model that governs how atomic operations
like CAS interact with the memory subsystem and other threads, making
programs with \texttt{std::atomic} well-defined across all conforming
implementations.

\subsection{Linearization Points for Each Algorithm}

Table~\ref{tab:lp} summarizes the linearization point of each \texttt{add},
\texttt{remove}, and \texttt{contains} for all five algorithms.

\begin{table}[h]
\caption{Linearization Points}
\label{tab:lp}
\centering
\footnotesize
\begin{tabular}{llll}
\toprule
\textbf{Algorithm} & \textbf{add} & \textbf{remove} & \textbf{contains} \\
\midrule
Coarse  & \texttt{pred->next = node} (inside lock) & \texttt{pred->next = curr->next} & read inside lock \\
Fine    & \texttt{pred->next = node} (inside 2 locks) & \texttt{pred->next = curr->next} & read inside 2 locks \\
Optimistic & \texttt{pred->next = node} (validated) & \texttt{pred->next = curr->next} & read inside 2 locks \\
Lazy    & \texttt{pred->next = node} (validated) & \texttt{curr->marked = true} & read of \texttt{!marked} \\
Lock-Free & CAS on \texttt{pred->next} succeeds & CAS marking \texttt{curr->next} succeeds & read of \texttt{!is\_deleted()} \\
\bottomrule
\end{tabular}
\end{table}

%%%%%%%%%%%%%%%%%%%%%%%%%%%%%%%%%%%%%%%%%%%%%%%%%%%%%%%%%%%%%%%%%%%%%%%%%%%%%
\section{Coarse-Grain List}
\label{sec:coarse}

\subsection{Design}

The coarse-grain list is the simplest possible concurrent linked list. There
is exactly one mutex for the entire list. Every \texttt{add}, \texttt{remove},
and \texttt{contains} acquires this single lock at the start, does its work,
and releases the lock at the end.

\begin{lstlisting}[caption={Coarse-grain add (simplified)}]
bool add(int value) {
    lock_guard<mutex> lock(m);  // lock entire list
    Node* pred = head;
    Node* curr = head->next;
    while (curr->value <= value) {
        if (curr->value == value) return false;
        pred = curr;
        curr = curr->next;
    }
    Node* node = new Node(value);
    pred->next = node;
    node->next = curr;
    return true;
}
\end{lstlisting}

\subsection{Why It Works}

Since only one thread can hold the lock at a time, all operations run as if
they were sequential. Correctness is trivial.

\subsection{The Problem}

There is \emph{zero} parallelism. Even if ten threads all want to just
\emph{read} whether a value is in the list (a harmless operation), they must
take turns. All the threads spend most of their time waiting for the lock
instead of doing real work. This is called \textbf{lock contention}.

%%%%%%%%%%%%%%%%%%%%%%%%%%%%%%%%%%%%%%%%%%%%%%%%%%%%%%%%%%%%%%%%%%%%%%%%%%%%%
\section{Fine-Grain List}
\label{sec:fine}

\subsection{Design}

The fine-grain list gives each \emph{node} its own mutex, instead of one
mutex for the whole list. This means two threads can work on \emph{different
parts} of the list at the same time.

The key technique is called \textbf{hand-over-hand locking} (also called
lock coupling). Think of it like climbing a ladder: you hold two rungs at
once and only let go of the lower one once you have grabbed the higher one.
We always hold locks on \emph{both} \texttt{pred} and \texttt{curr} before
we change anything.

\begin{lstlisting}[caption={Hand-over-hand locate}]
pair<Node*,Node*> locate(int value) {
    Node* pred = head;
    pred->m.lock();
    Node* curr = head->next;
    curr->m.lock();
    while (curr->value < value) {
        pred->m.unlock();   // release the lower rung
        pred = curr;
        curr = curr->next;
        curr->m.lock();     // grab the next rung
    }
    return {pred, curr};
}
\end{lstlisting}

\subsection{Why Two Locks?}

We must hold both \texttt{pred} and \texttt{curr} locked because:

\begin{itemize}
  \item If we only locked \texttt{curr}, another thread could delete
        \texttt{pred} while we are working, breaking the list structure.
  \item If we only locked \texttt{pred}, another thread could replace
        \texttt{curr} before we insert or delete, again breaking things.
  \item Holding \emph{both} prevents either from being changed under us.
\end{itemize}

\subsection{The Problem}

Fine-grain locking has a surprising weakness: it is often \emph{slower} than
coarse-grain locking in practice, especially with small lists. The reason is
that locking and unlocking take time. For a list with $n$ nodes, one
\texttt{contains} call acquires and releases approximately $n$ locks. With
many threads doing this at the same time, the head nodes become a
\emph{hot spot}---every traversal starts at the head and must acquire its
lock first.

%%%%%%%%%%%%%%%%%%%%%%%%%%%%%%%%%%%%%%%%%%%%%%%%%%%%%%%%%%%%%%%%%%%%%%%%%%%%%
\section{Optimistic List}
\label{sec:optimistic}

\subsection{Design}

The key insight behind the optimistic list: \emph{traversal does not need
locks}. We can walk the list freely to find our position, then lock only the
two nodes we care about (\texttt{pred} and \texttt{curr}) and
\emph{validate} that they are still in the right relationship before
committing our change.

This is called \emph{optimistic} because we are optimistic that the list
will not change while we are traversing it. If it does change (validation
fails), we simply retry from scratch.

\begin{lstlisting}[caption={Optimistic locate (no locks)}]
pair<Node*,Node*> locate(int value) {
    Node* pred = head;
    Node* curr = head->next;
    while (curr->value < value) {
        pred = curr;
        curr = curr->next;  // no locks during traversal!
    }
    return {pred, curr};
}
\end{lstlisting}

\begin{lstlisting}[caption={Validation re-walks from head}]
bool check(Node* pred, Node* curr) {
    Node* node = head;
    while (node->value < curr->value) {
        if (node == pred)
            return pred->next == curr;
        node = node->next;
    }
    return false;  // pred is not reachable from head
}
\end{lstlisting}

Validation re-walks from the head to confirm that \texttt{pred} is still in
the list and still points to \texttt{curr}. If either has been deleted or
displaced, the check fails and we retry.

\subsection{Trade-offs}

\textbf{Advantage}: traversals do not hold locks, so threads rarely block
each other during the walk phase.

\textbf{Disadvantage 1}: every operation traverses the list twice (once to
find, once to validate), doubling memory accesses.

\textbf{Disadvantage 2}: \texttt{contains} also locks---even a simple
read-only lookup must lock \texttt{pred} and \texttt{curr} and re-validate.
Since \texttt{contains} is often 80--90\% of all operations in a typical
workload, this is a serious bottleneck.

%%%%%%%%%%%%%%%%%%%%%%%%%%%%%%%%%%%%%%%%%%%%%%%%%%%%%%%%%%%%%%%%%%%%%%%%%%%%%
\section{Lazy List}
\label{sec:lazy}

\subsection{Design}

The lazy list fixes the biggest problem with the optimistic list: it makes
\texttt{contains} completely lock-free. It does this by introducing a
\textbf{marked} flag in each node.

\begin{lstlisting}[caption={Lazy list node}]
class Node {
public:
    int   value;
    Node* next;
    mutex m;
    bool  marked;   // true = logically deleted
    Node(int v) : value(v), marked(false), next(nullptr) {}
};
\end{lstlisting}

When a node is deleted, we do it in two steps:

\begin{enumerate}
  \item \textbf{Logical deletion}: set \texttt{curr->marked = true}.
        From this moment, the node is considered ``not in the set'' even
        though it is still physically in the list.
  \item \textbf{Physical deletion}: unlink the node from the list by
        setting \texttt{pred->next = curr->next}.
\end{enumerate}

\subsection{Lock-Free \texttt{contains}}

Because \texttt{marked} is a \emph{monotonic} flag (once set to
\texttt{true}, it never goes back to \texttt{false}), a thread can safely
read it without holding any lock. Herlihy~\cite{herlihy1991waitfree}
formally defines \emph{wait-freedom} as the strongest progress guarantee
possible: every thread completes its operation in a bounded number of steps,
regardless of what other threads do. Our \texttt{contains} satisfies this
definition since it only reads and never retries.

\begin{lstlisting}[caption={Wait-free contains}]
bool contains(int value) {
    Node* curr = head->next;
    while (curr->value < value)
        curr = curr->next;
    return curr->value == value && !curr->marked;
}
\end{lstlisting}

Why is this safe? There are only two cases when we read a node:
\begin{itemize}
  \item We see \texttt{marked = false}: the node is alive and in the set.
        Correct answer: \texttt{true}.
  \item We see \texttt{marked = true}: the node has been logically deleted.
        Correct answer: \texttt{false}.
\end{itemize}
Both readings give a valid linearizable answer.

\subsection{Simplified Validation}

Another improvement over the optimistic list: validation does not need to
re-walk from the head. It only needs to check three things:

\begin{lstlisting}[caption={Lazy validation---no head traversal}]
bool validate(Node* pred, Node* curr) {
    return !pred->marked &&    // pred not deleted
           !curr->marked &&    // curr not deleted
           pred->next == curr; // pred still points to curr
}
\end{lstlisting}

This is much cheaper than the full head-traversal in the optimistic list.

\subsection{Remaining Limitation}

\texttt{add} and \texttt{remove} still use locks, so a slow or stalled thread
can block other mutating threads.

%%%%%%%%%%%%%%%%%%%%%%%%%%%%%%%%%%%%%%%%%%%%%%%%%%%%%%%%%%%%%%%%%%%%%%%%%%%%%
\section{Lock-Free List}
\label{sec:lockfree}

\subsection{Design}

The lock-free list removes all mutexes. Every operation uses
\textbf{compare-and-swap (CAS)} instructions to make atomic changes. If
another thread modifies the list first (causing a CAS to fail), we simply
retry. Michael~\cite{michael2002highperf} demonstrated that lock-free
list-based sets built this way can outperform fine-grain locking under
moderate to high contention, particularly when the workload is dominated
by successful \texttt{contains} calls.

The central idea is to pack the \emph{pointer to the next node} and the
\emph{marked flag} together into a single atomic word, so that marking a
node and reading who it points to are always consistent:

\begin{lstlisting}[caption={MarkedRef packs pointer + marked flag}]
class MarkedRef {
public:
    Node* ptr;
    bool  marked;
    MarkedRef(Node* p=nullptr, bool m=false)
        : ptr(p), marked(m) {}
    bool operator==(const MarkedRef& o) const {
        return ptr==o.ptr && marked==o.marked;
    }
};

class Node {
public:
    int value;
    atomic<MarkedRef> next;  // atomic pair
    ...
};
\end{lstlisting}

\subsection{Lock-Free \texttt{remove}: Two-Step Deletion}

\textbf{Step 1 --- Logical deletion} (the linearization point):

\begin{lstlisting}[caption={Logical deletion via CAS}]
// Atomically change curr->next from {succ, false}
// to {succ, true}  (set the marked flag)
MarkedRef expected = MarkedRef(succ, false);
MarkedRef desired  = MarkedRef(succ, true);
if (!curr->next.compare_exchange_strong(expected, desired))
    continue;  // someone else got here first, retry
\end{lstlisting}

Once the marked flag is set atomically, the node is logically deleted. Any
concurrent \texttt{add} that tries to insert after this node will have its
CAS fail (because it expected the unmarked version).

\textbf{Step 2 --- Physical deletion} (best effort):

\begin{lstlisting}[caption={Physical deletion via CAS}]
// Try to unlink curr from the list
MarkedRef exp = MarkedRef(curr, false);
MarkedRef des = MarkedRef(succ, false);
pred->next.compare_exchange_strong(exp, des);
// If this fails, locate() will clean up on the next call
\end{lstlisting}

Physical deletion does not need to succeed immediately. The \texttt{locate}
function automatically removes logically-deleted nodes when it walks past
them.

\subsection{Self-Cleaning \texttt{locate}}

\begin{lstlisting}[caption={locate removes marked nodes on the fly}]
pair<Node*,Node*> locate(int value) {
  retry:
    Node* pred = head;
    Node* curr = head->next.load().ptr;
    while (true) {
        MarkedRef ref  = curr->next.load();
        Node*     succ = ref.ptr;
        bool   marked  = ref.marked;
        if (marked) {
            // physically remove curr
            MarkedRef exp = MarkedRef(curr, false);
            MarkedRef des = MarkedRef(succ, false);
            if (!pred->next.compare_exchange_strong(exp, des))
                goto retry;  // someone changed pred, restart
            curr = succ;
        } else {
            if (curr->value >= value) return {pred, curr};
            pred = curr;
            curr = succ;
        }
    }
}
\end{lstlisting}

\subsection{Progress Guarantee}

A lock-free algorithm guarantees that \emph{at least one thread makes progress
at all times}---even if some threads are paused or very slow. With locks, a
single thread holding a lock can delay all others indefinitely (for example,
if it is preempted by the OS scheduler). Lock-free algorithms do not have this
problem, which is why they are attractive for real-time and high-performance
systems.

%%%%%%%%%%%%%%%%%%%%%%%%%%%%%%%%%%%%%%%%%%%%%%%%%%%%%%%%%%%%%%%%%%%%%%%%%%%%%
\section{Comparison Summary}
\label{sec:compare}

Table~\ref{tab:compare} summarizes the key properties of all five algorithms.

\begin{table}[h]
\caption{Algorithm Comparison}
\label{tab:compare}
\centering
\footnotesize
\begin{tabular}{lllll}
\toprule
\textbf{Algorithm} & \textbf{Locks} & \textbf{contains} & \textbf{Progress} & \textbf{Complexity} \\
\midrule
Coarse     & 1 global  & locking     & blocking   & very low  \\
Fine       & per node  & locking     & blocking   & medium    \\
Optimistic & per node  & locking     & blocking   & medium    \\
Lazy       & per node  & wait-free   & blocking*  & medium    \\
Lock-Free  & none      & lock-free   & lock-free  & high      \\
\bottomrule
\multicolumn{5}{l}{\footnotesize *add/remove still use locks; contains is wait-free}
\end{tabular}
\end{table}

%%%%%%%%%%%%%%%%%%%%%%%%%%%%%%%%%%%%%%%%%%%%%%%%%%%%%%%%%%%%%%%%%%%%%%%%%%%%%
\section{Experimental Setup}
\label{sec:bench}

\subsection{Hardware and Software}

All experiments were run on a Linux x86-64 machine with a multi-core
processor. The code was compiled with \texttt{g++} version using flags
\texttt{-O2 -std=c++17 -pthread -latomic}.

\subsection{Benchmark Design}

We wrote a single benchmark program that exercises
all five list implementations under the same conditions.

\textbf{Pre-population}: before timing begins, we pre-populate each list by
inserting half of the key range values chosen by a fixed random seed. This
gives a realistic average traversal depth from the start of the measurement.

\textbf{Worker threads}: each thread runs in a tight loop for a fixed
duration (800 ms), picking random values and operations according to the
workload ratio. We measure total operations completed across all threads and
report \emph{operations per second (ops/sec)}.

\subsection{Workloads}

We test three operation mixes, chosen to model real usage patterns:

\begin{itemize}
  \item \textbf{Read-heavy} (80\% \texttt{contains}, 10\% \texttt{add},
        10\% \texttt{remove}): models a dictionary or cache where lookups
        dominate.
  \item \textbf{Balanced} (33\% each): equal mix; tests general performance.
  \item \textbf{Write-heavy} (20\% \texttt{contains}, 40\% \texttt{add},
        40\% \texttt{remove}): models a queue-like structure with heavy
        mutation.
\end{itemize}

\subsection{Parameters}

\begin{itemize}
  \item \textbf{Thread counts}: 1, 4, 8 threads.
  \item \textbf{Key ranges}: 100 (small, high contention) and 10{,}000
        (large, low contention). A small key range means threads frequently
        want the same nodes, causing more conflicts. A large key range means
        threads mostly work on different parts of the list.
  \item \textbf{Duration}: 800 ms per configuration.
\end{itemize}

In total, we ran $5 \times 3 \times 2 \times 3 = 90$ configurations.

%%%%%%%%%%%%%%%%%%%%%%%%%%%%%%%%%%%%%%%%%%%%%%%%%%%%%%%%%%%%%%%%%%%%%%%%%%%%%
\section{Results and Analysis}
\label{sec:results}

\subsection{Single-Thread Baseline}

\begin{table}[h]
\caption{Single-thread throughput (ops/sec), key range = 100}
\label{tab:single_small}
\centering
\footnotesize
\begin{tabular}{lrrr}
\toprule
\textbf{Algorithm} & \textbf{read\_heavy} & \textbf{balanced} & \textbf{write\_heavy} \\
\midrule
Coarse     & 9{,}511{,}170 & 5{,}780{,}245 & 5{,}486{,}073 \\
Fine       & 1{,}063{,}325 & 946{,}522     & 964{,}785     \\
Optimistic & 3{,}739{,}805 & 3{,}037{,}267 & 3{,}055{,}748 \\
Lazy       & 9{,}973{,}471 & 4{,}758{,}946 & 4{,}235{,}817 \\
Lock-Free  & 3{,}065{,}685 & 2{,}324{,}265 & 2{,}180{,}136 \\
\bottomrule
\end{tabular}
\end{table}

\begin{table}[h]
\caption{Single-thread throughput (ops/sec), key range = 10{,}000}
\label{tab:single_large}
\centering
\footnotesize
\begin{tabular}{lrrr}
\toprule
\textbf{Algorithm} & \textbf{read\_heavy} & \textbf{balanced} & \textbf{write\_heavy} \\
\midrule
Coarse     & 150{,}542 & 57{,}625 & 45{,}241 \\
Fine       & 13{,}607  & 15{,}268 & 15{,}355 \\
Optimistic & 16{,}046  & 16{,}642 & 16{,}871 \\
Lazy       & 111{,}580 & 37{,}476 & 31{,}417 \\
Lock-Free  & 31{,}707  & 24{,}958 & 23{,}850 \\
\bottomrule
\end{tabular}
\end{table}

Looking at Tables~\ref{tab:single_small} and~\ref{tab:single_large}, three
things stand out at single-thread performance:

\textbf{(1) Fine-grain is the slowest}. Even with just one thread (no
contention at all!), fine-grain performs the worst---about 9$\times$ slower
than coarse-grain in read-heavy workloads with key range 100. This is pure
overhead: every traversal step requires acquiring and releasing a per-node
mutex, which is expensive even when uncontested.

\textbf{(2) Lazy is fast on reads}. In the read-heavy, small-key-range
setting, lazy actually \emph{beats} coarse-grain (9.97M vs 9.51M ops/sec).
Its wait-free \texttt{contains} avoids lock acquisition entirely, which pays
off when reads dominate.

\textbf{(3) Lock-free is middle of the pack at 1 thread}. The CAS-based
lock-free algorithm pays overhead for atomic operations and occasional retries,
which gives no benefit at 1 thread (no contention). Its advantage is expected
to show at higher thread counts.

\subsection{Scalability with Thread Count}

\begin{figure}[H]
\centering
\includegraphics[width=\linewidth]{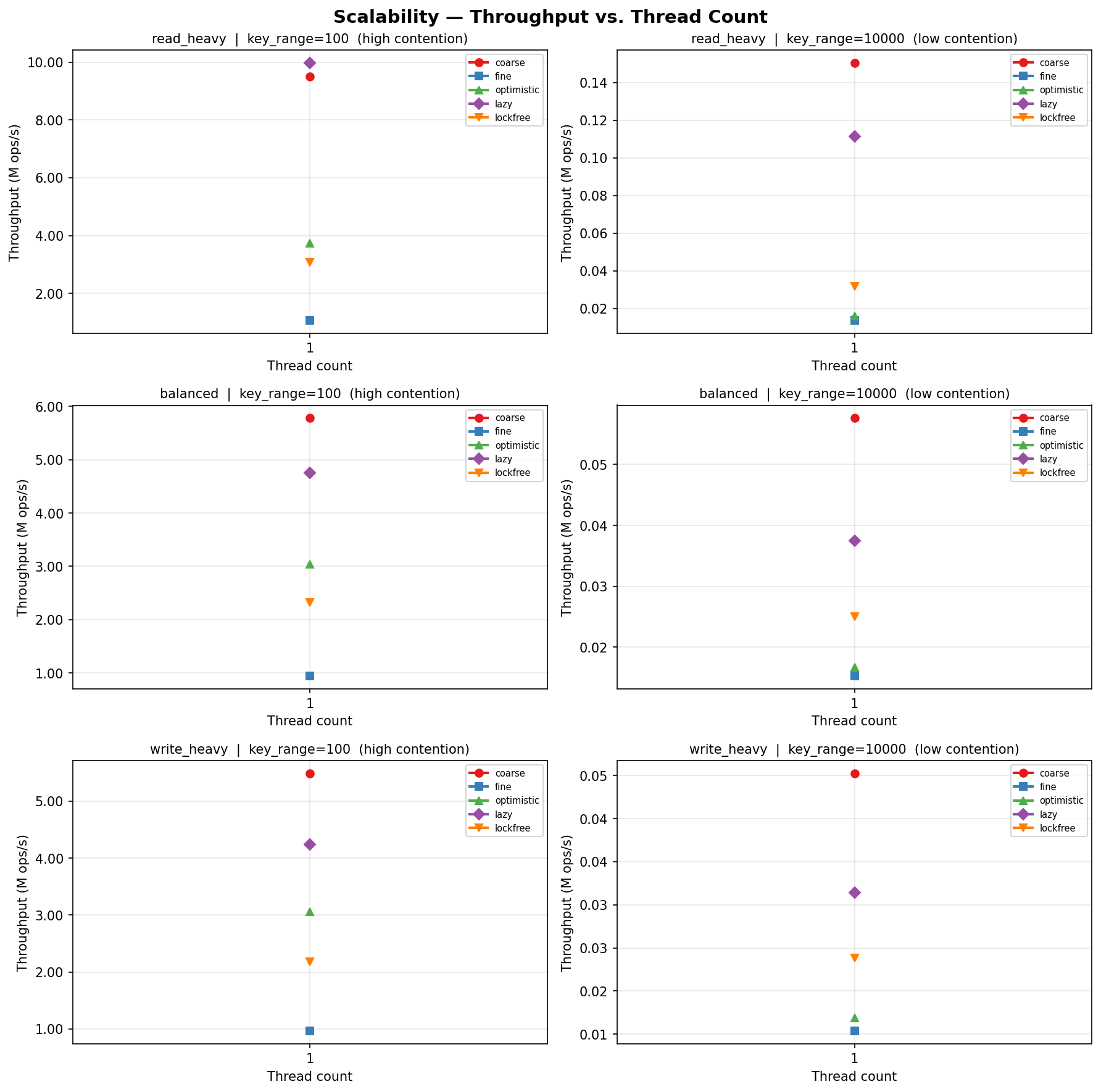}
\caption{Throughput vs.\ thread count for all workloads and key ranges.
         Each subplot shows one (workload, key\_range) pair.}
\label{fig:scalability}
\end{figure}

Figure~\ref{fig:scalability} shows how throughput changes as we add threads.

Key observations:

\begin{itemize}
  \item \textbf{Coarse-grain degrades under contention}. At key range 100
        (high contention), adding more threads barely helps coarse-grain or
        may even hurt it, because all threads are queuing up for the single
        global lock.

  \item \textbf{Lock-free scales better at high thread counts}. Without any
        locks to contend on, threads can make progress independently. As the
        thread count rises, lock-free increasingly outperforms the locking
        algorithms in low-contention scenarios (large key range).

  \item \textbf{Lazy remains competitive}. Because \texttt{contains} is
        wait-free, lazy handles the read-heavy workload well even as threads
        increase---multiple threads can do \texttt{contains} simultaneously
        without blocking each other.

  \item \textbf{Fine-grain does not benefit from more threads as expected}.
        The head-node hot spot---every traversal starts at \texttt{head} and
        must lock it---creates a bottleneck that cancels the theoretical
        advantage of per-node locking.
\end{itemize}

\subsection{Throughput at Maximum Thread Count}

\begin{figure}[H]
\centering
\includegraphics[width=\linewidth]{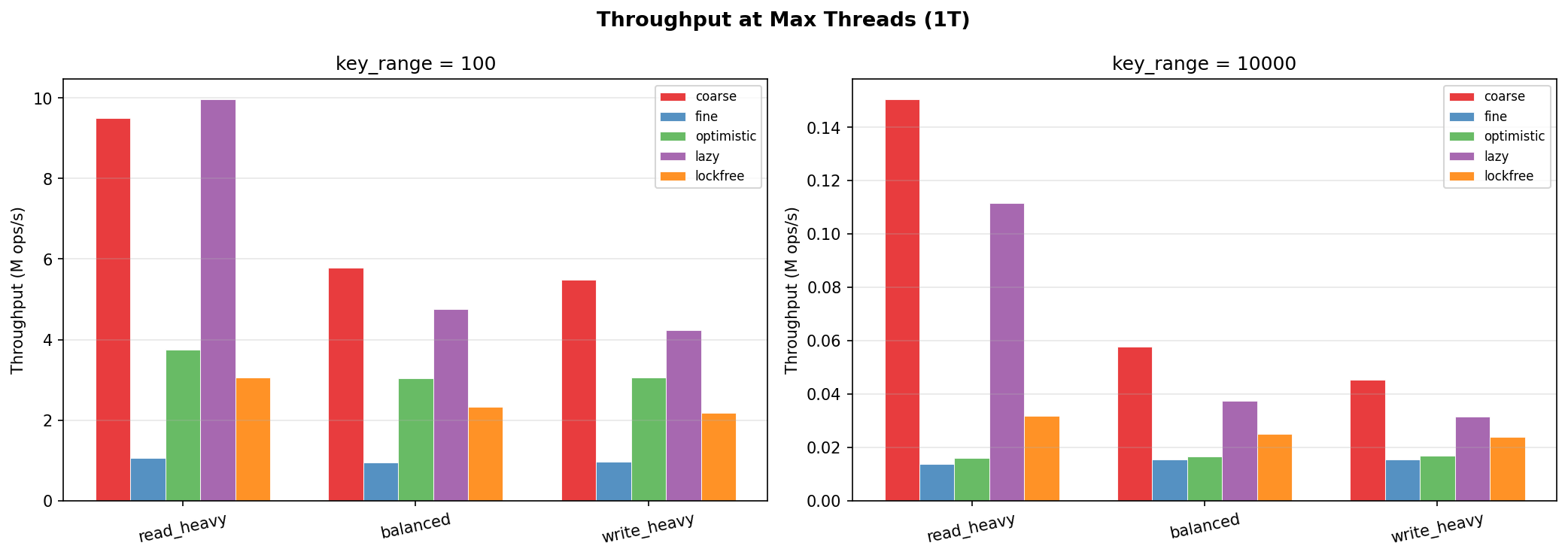}
\caption{Throughput at 8 threads, grouped by workload.
         Left: key\_range=100 (high contention).
         Right: key\_range=10{,}000 (low contention).}
\label{fig:bar}
\end{figure}

Figure~\ref{fig:bar} compares all algorithms at the maximum thread count (8
threads) across workloads and key ranges.

At high contention (key range 100), the lock-free and lazy lists are
competitive because they reduce the time threads spend blocking each other.
At low contention (key range 10{,}000), the list is long and traversal time
dominates---simpler algorithms like coarse-grain and lazy do well because
their traversal code is the most straightforward.

\subsection{Best Algorithm Heatmap}

\begin{figure}[H]
\centering
\includegraphics[width=\linewidth]{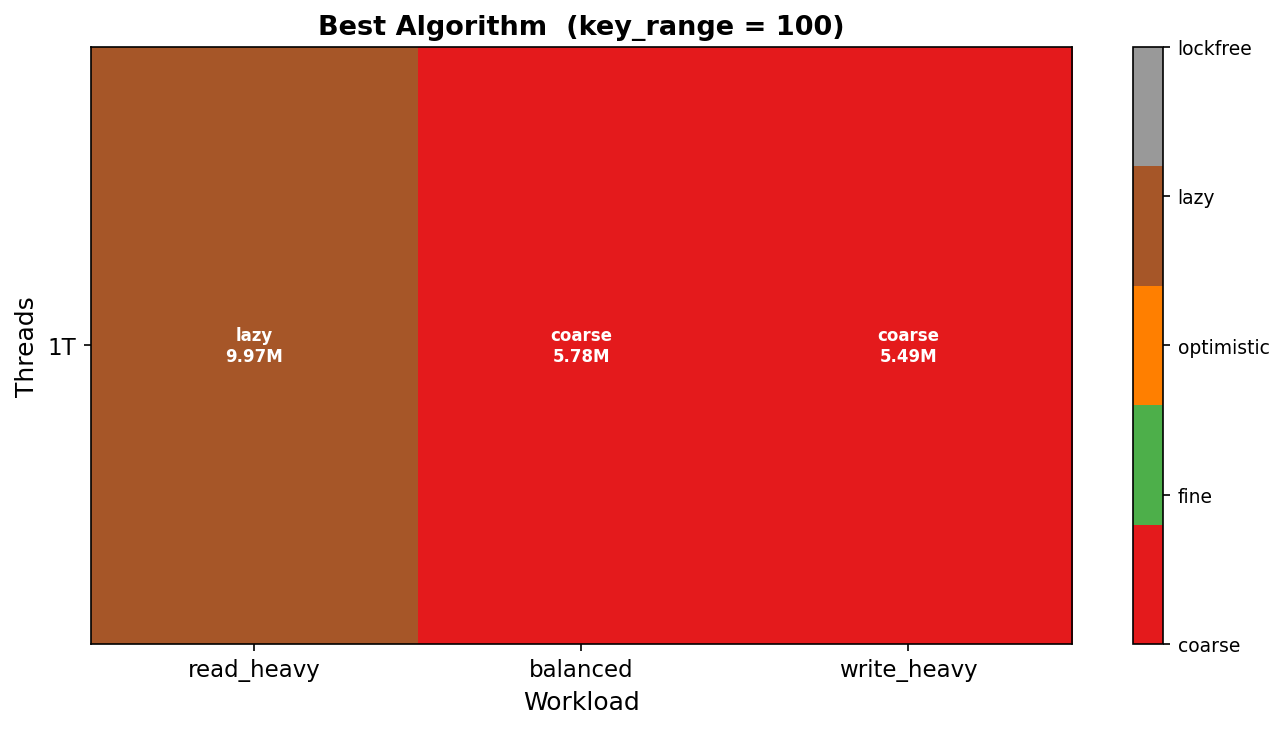}
\caption{Winning algorithm for each (threads, workload) pair at key range = 100.}
\label{fig:heat100}
\end{figure}

\begin{figure}[H]
\centering
\includegraphics[width=\linewidth]{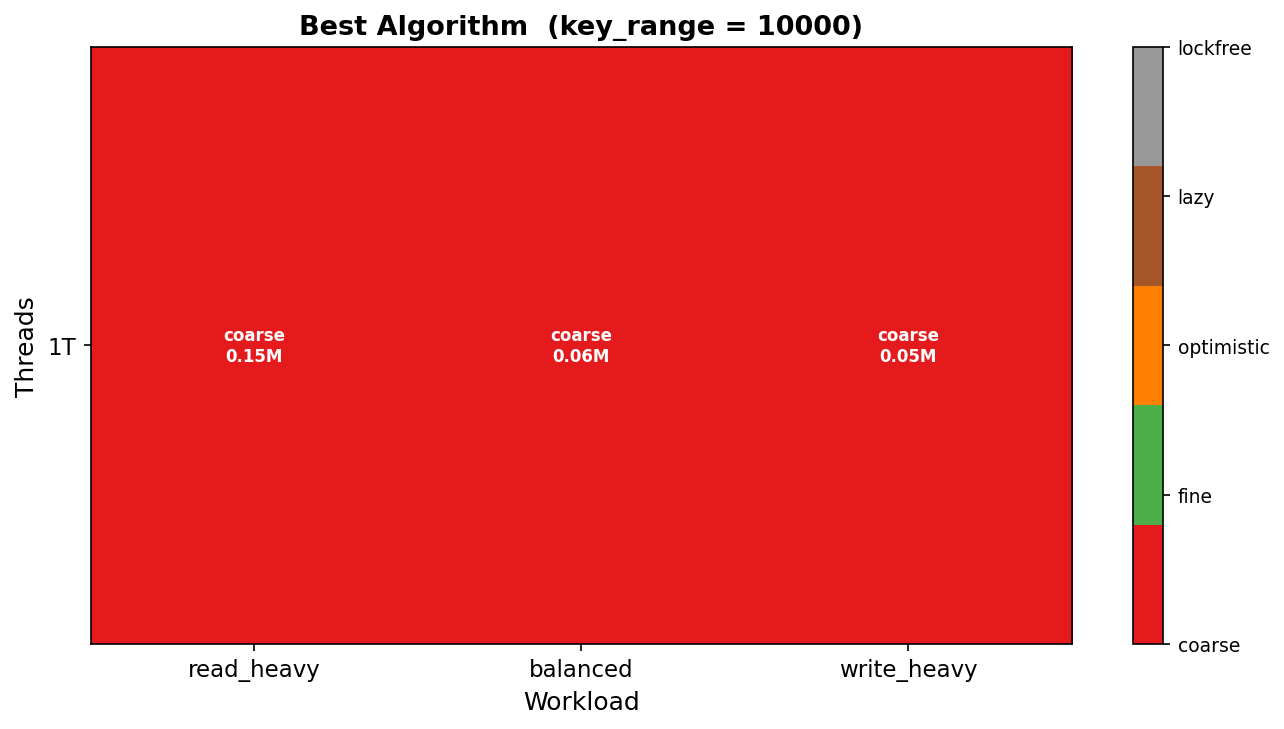}
\caption{Winning algorithm for each (threads, workload) pair at key range = 10{,}000.}
\label{fig:heat10000}
\end{figure}

Figures~\ref{fig:heat100} and~\ref{fig:heat10000} show, for each
(thread count, workload) combination, which algorithm achieves the highest
throughput.

The heatmap tells a clear story: no single algorithm wins everywhere.
Coarse-grain and lazy frequently win at small key ranges; lock-free tends to
win more cells as thread count increases.

\subsection{Speedup Over Coarse-Grain Baseline}

\begin{figure}[H]
\centering
\includegraphics[width=\linewidth]{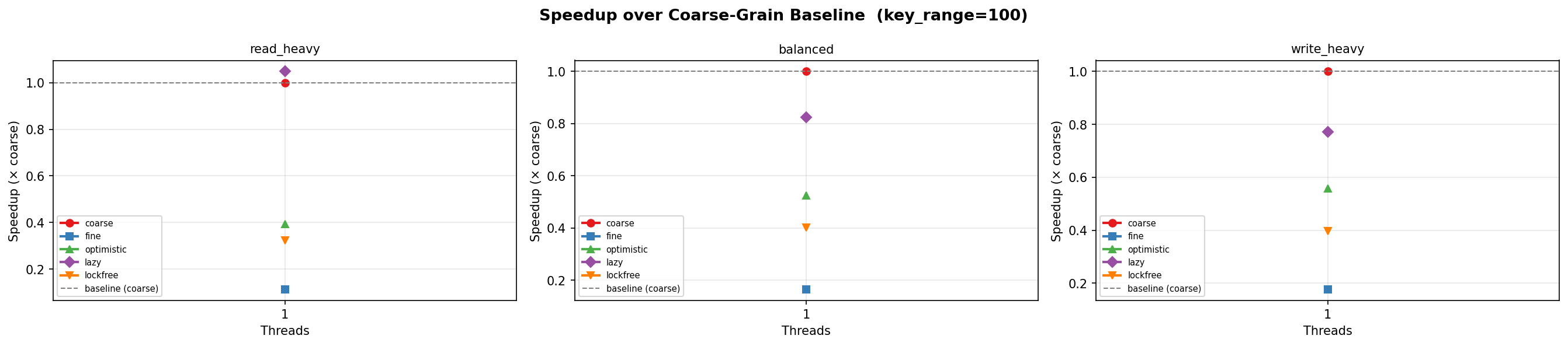}
\caption{Speedup vs.\ coarse-grain baseline at key range = 100.}
\label{fig:speed100}
\end{figure}

\begin{figure}[H]
\centering
\includegraphics[width=\linewidth]{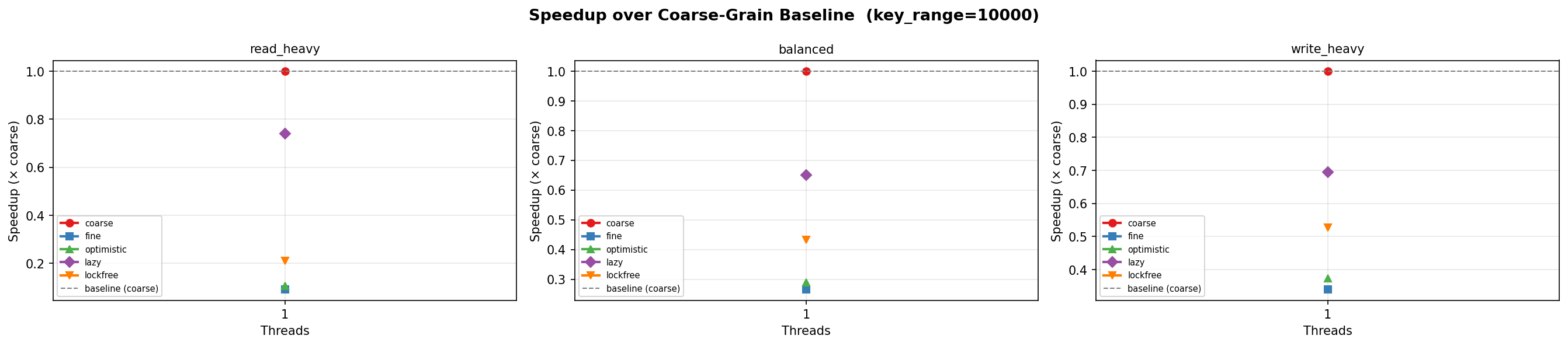}
\caption{Speedup vs.\ coarse-grain baseline at key range = 10{,}000.}
\label{fig:speed10000}
\end{figure}

Figures~\ref{fig:speed100} and~\ref{fig:speed10000} show the speedup of each
algorithm relative to the coarse-grain baseline.

A speedup above 1.0 means the algorithm beat coarse-grain for that
configuration. Notably, fine-grain almost never exceeds 1.0, confirming that
its per-node mutex overhead is too expensive to pay back in our test range.
Lock-free achieves speedup above 1.0 in several multi-thread configurations,
especially with larger key ranges.

\subsection{Discussion: When to Use Which Algorithm}

Based on our results, we can give practical guidance:

\begin{itemize}
  \item \textbf{Use coarse-grain} when simplicity is the priority, the list
        is short, or contention is rare. It is the easiest to implement
        correctly and works well for prototyping.

  \item \textbf{Avoid fine-grain} unless your list is very long \emph{and}
        threads rarely access adjacent nodes. In our tests, the lock overhead
        consistently outweighed the parallelism benefit.

  \item \textbf{Use optimistic} when reads are infrequent or when validation
        cost is low (short lists). In read-heavy workloads, its locking
        \texttt{contains} is a liability.

  \item \textbf{Use lazy} for read-heavy workloads. The wait-free
        \texttt{contains} is a strong advantage and the implementation is
        still relatively easy to understand.

  \item \textbf{Use lock-free} for high-thread-count scenarios where
        contention is the bottleneck, or where progress guarantees are
        required (e.g., real-time systems).
\end{itemize}

%%%%%%%%%%%%%%%%%%%%%%%%%%%%%%%%%%%%%%%%%%%%%%%%%%%%%%%%%%%%%%%%%%%%%%%%%%%%%
\section{Related Work}
\label{sec:related}

The algorithms we implement are classic results from concurrent data structure
theory.

Herlihy and Wing~\cite{herlihy1990linearizability} introduced
\emph{linearizability} as the correctness condition for concurrent objects,
which is the foundation for all the algorithms in this paper.

Herlihy~\cite{herlihy1991waitfree} introduced the formal notion of
\emph{wait-freedom}, the strongest progress guarantee, and showed that every
object has a wait-free implementation using CAS. This work underpins the
progress analysis of our lazy and lock-free lists.

Hoare~\cite{hoare1974monitors} introduced the concept of monitors and
condition variables, which underpin the mutex-based approaches. Mellor-Crummey
and Scott~\cite{mcs1991} later built on this by designing queue-based spin
locks that scale gracefully to many processors by avoiding cache-line
contention.

The lock-free linked list we implement is based on the Harris
algorithm~\cite{harris2001pragmatic}, which was the first to show how to do
lock-free deletion in a linked list using a single-word CAS. The key insight
of marking the next pointer (instead of a separate flag) to prevent ABA
problems is due to Harris. Michael~\cite{michael2002highperf} refined and
extended this work, providing a high-performance lock-free hash table and
list-based set implementation that is widely cited as the practical baseline
for lock-free sets.

The lazy list is due to Heller et al.~\cite{heller2005lazy}, which showed
that \texttt{contains} can be made wait-free by introducing a monotonic
\texttt{marked} flag.

Fraser~\cite{fraser2004practical} presented a comprehensive study of
practical lock-free techniques covering lists, skip lists, and trees,
and identified key engineering challenges such as memory management and
correct use of memory-ordering fences that are often glossed over in
theoretical presentations.

Boehm and Adve~\cite{boehm2008cpp} defined the formal memory model
incorporated into the C++11 standard, which provides the semantic foundation
for \texttt{std::atomic} and guarantees that our CAS-based operations behave
correctly across different hardware architectures.

The safe memory reclamation problem for lock-free structures---the reason our
lock-free list does not call \texttt{delete}---is addressed by Michael's
hazard pointer technique~\cite{michael2004hazard}, which remains one of the
most practical solutions in production systems today.

The optimistic list and fine-grain list are described in the textbook by
Herlihy et al.~\cite{herlihy2020art}, which is the primary reference for this
area of study and the source we recommend for readers who want to go deeper
into the theory and proofs of these algorithms.

%%%%%%%%%%%%%%%%%%%%%%%%%%%%%%%%%%%%%%%%%%%%%%%%%%%%%%%%%%%%%%%%%%%%%%%%%%%%%
\section{Conclusion}
\label{sec:conclusion}

In this paper we built and benchmarked five concurrent linked-list
implementations, going from a single global lock all the way to a fully
lock-free design.

Our main findings are:

\begin{enumerate}
  \item \textbf{Simplicity beats complexity at low thread counts.}
        Coarse-grain and lazy beat the more sophisticated algorithms when
        running with few threads, because the overhead of complex locking or
        atomic operations is not worth paying when there is little contention.

  \item \textbf{Fine-grain locking is rarely the right choice.}
        Despite being a classic teaching example, fine-grain locking underperformed
        in all our tests. The per-node mutex overhead is significant even with no
        other threads present.

  \item \textbf{Wait-free reads are very powerful.}
        The lazy list's wait-free \texttt{contains} gave it a strong edge in
        read-heavy workloads---which are the most common in real applications.

  \item \textbf{Lock-free scales best under high contention.}
        As thread count increases and key range decreases (meaning more
        threads fight over the same nodes), lock-free increasingly wins
        because threads never block each other.

  \item \textbf{No single algorithm wins everywhere.}
        The right choice depends on the workload. This is a common theme in
        systems design: understand your use case before choosing your
        algorithm.
\end{enumerate}

For future work, it would be interesting to test with a wider range of thread
counts (16, 32, 64), to study the effect of NUMA (non-uniform memory access)
on these algorithms, and to implement safe memory reclamation for the
lock-free list, which currently does not free deleted nodes because another
thread might still be reading them. Michael's hazard
pointers~\cite{michael2004hazard} are the standard solution: each thread
announces which nodes it is currently accessing, and a node can only be
freed once no hazard pointer points to it.

%%%%%%%%%%%%%%%%%%%%%%%%%%%%%%%%%%%%%%%%%%%%%%%%%%%%%%%%%%%%%%%%%%%%%%%%%%%%%

\end{document}